\documentclass[useAMS,usenatbib]{mn2e}
\usepackage[english]{babel}
\usepackage{graphicx}
\usepackage{amsmath}
\usepackage{amssymb}
\usepackage{BibDef}

\usepackage{subfigure}
\usepackage{epstopdf}

\def\msun{{\rm M}_\odot}

\def\lsim{\mathrel{\rlap{\lower 3pt \hbox{$\sim$}} \raise 2.0pt \hbox{$<$}}}
\def\gsim{\mathrel{\rlap{\lower 3pt \hbox{$\sim$}} \raise 2.0pt \hbox{$>$}}}

\title[Not bars but bar formation]{Bar-formation as driver of gas inflows in
isolated disc galaxies}
                    
\author[R. Fanali et al.]
{R.~Fanali$^{1}$, M.~Dotti$^{1,2}$, D.~Fiacconi$^{3}$ \& F.~Haardt$^{4,2}$
\\
$^1$ Universit\`a degli Studi di Milano Bicocca, I--20126 Milano, Italy\\
$^2$ INFN, Sezione di Milano--Bicocca, Piazza della Scienza 3, I--20126 Milano, Italy\\
$^3$ Institute for Computational Science, University of Z\"urich,
Winterthurerstrasse 190, CH-8057 Z\"urich, Switzerland \\
$^4$ DiSAT, Universit\`a dell'Insubria, via Valleggio 11, I--22100 Como, Italy \\
}

\begin{document}

\date{\today}

\pagerange{\pageref{firstpage}--\pageref{lastpage}} \pubyear{2015}

\maketitle

\label{firstpage}


\begin{abstract}
Stellar bars are a common feature in massive disc galaxies. On a
theoretical ground, the response of gas to a bar is generally thought
to cause nuclear starbursts and, possibly, AGN activity once the
perturbed gas reaches the central super-massive black hole. By means
of high resolution numerical simulations we detail the purely
dynamical effects that a forming bar exerts on the gas of an isolated
disc galaxy. The galaxy is initially unstable to the formation of
non-axisymmetric structures, and within $\sim 1$ Gyr it develops
spiral arms that eventually evolve into a central stellar bar on kpc
scale.  A first major episode of gas inflow occurs during the
formation of the spiral arms while at later times, when the stellar
bar is establishing, a low density region is carved between the bar
co-rotational and inner Lindblad resonance radii. The development of
such ``dead zone" inhibits further massive gas inflows. Indeed, the
gas inflow reaches its maximum during the relatively fast bar
formation phase and not, as often assumed, when the bar is fully
formed. We conclude that the low efficiency of long-lived, evolved
bars in driving gas toward galactic nuclei is the reason why
observational studies have failed to establish an indisputable link
between bars and AGNs.  On the other hand, the high efficiency in
driving strong gas inflows of the intrinsically transient process of
bar formation suggests that the importance of bars as drivers of AGN
activity in disc galaxies has been overlooked so far.  We finally
prove that our conclusions are robust against different numerical
implementations of the hydrodynamics routinely used in galaxy
evolution studies.
\end{abstract}



\section{Introduction}

The fraction of disc galaxies showing a well developed stellar bar in the
local Universe is substantial, up to $\gsim 30\%$ for massive ($M_* \gsim
10^{9.5} \msun$) systems \citep{Laurikainen04,Nair10,Lee12a,Gavazzi15}.  The
effectiveness of bars in modifying
the dynamics of gas has been recognized since decades
\citep[e.g.][]{Sanders76, Roberts79, Athanassoula92}.  In particular, gas
within the bar corotational radius ($R_{C}$, i.e. the radius at which the
angular velocity in the disc plane $\Omega(R)$ equals the bar pattern
precession speed $\Omega_b$) is driven toward the centre of the galaxy because
of the interaction with the bar itself. Early theoretical studies suggested
that such inflows could be responsible for nuclear starbursts and, if the gas
is able to reach the very central regions of the galaxy, AGN activity
\citep[e.g.][]{Shlosman89, Berentzen98}.

From the observational point of view the connection between bars and enhanced 
nuclear star formation has been extensively proved 
\citep[e.g.][]{Ho97, Martinet97, Hunt99, Laurikainen04, Jogee05}. 
The link between bars and AGN seems less clear: while barred galaxies 
host AGNs more frequently that their non-barred analogous 
\citep[making bars a good candidate for the triggering of nuclear activity,
e.g.][]{Laurikainen04, Oh12}, 
it is still matter of debate whether the presence of bars is one of the main drivers
of AGNs 
\citep[as suggested by, e.g.][]{Knapen00, Laine02, Alonso13} or not 
\citep[see e.g.][]{Ho97, Mulchaey97, Hunt99, Lee12b, Cisternas13, Cheung15}.

In order to have a comprehensive understanding of the gas dynamics in barred
galaxies many numerical studies have been put forward, including, for example
both 2- or 3-D simulations, and different schemes for the gas hydrodynamics
(smoothed particle hydrodynamics, SPH, vs grid codes). We consider
particularly meaningful to divide the different efforts in three main classes:
\begin{enumerate}
\item {\bf Isolated galaxies with analytical bars}
\citep[e.g.][]{Athanassoula92,Regan04,Kim12}. 
In this class of simulations (often restricted to a 2-D geometry) bars are
represented by analytical potentials 
that do not evolve in time (but for their rigid body rotation). 
These simulations, although quite idealized, allows for extremely high resolutions
and precise 
evolution of the gas dynamics.
\item {\bf Fully evolving isolated galaxies}
\citep[e.g.][]{Berentzen98,Berentzen07,VillaVargas10,Cole14}, 
where bars are modeled (as the rest of the galaxy) as evolving structures, that can
change their extents, 
rotational patterns, etc.
\item {\bf Cosmological simulations}
\citep[e.g.][]{RomanoDiaz08,Scannapieco12,Kraljic12,Goz14,Fiacconi2015}. 
In these simulations the galaxies form from cosmological perturbations, and are free
to acquire mass 
and angular momentum through large scale gas inflows and galaxy mergers. 
In this approach the initial conditions are not arbitrary, but, because of the large
boxes simulated 
(even in zoom-in runs), the spatial and mass resolution is usually significantly
coarser than in 
isolated simulations.
\end{enumerate}

Simulations of the first kind have confirmed the analytical prediction that,
in many galactic potentials, bar-driven gas inflows fail to reach the very
centre of the galaxy. The gas shocks around the outermost inner Lindblad
resonance (ILR) radius ($R_{\rm ILR}$) of the bar, defined by the equality
$\Omega(R)-\kappa(R)/2=\Omega_b$ where $\kappa$ is the epicyclic frequency,
i.e. the frequency of small radial oscillations. At $R_{\rm ILR}$ the gas
shocks, forming nuclear rings that are often observed as star forming regions
in barred galaxies \citep[e.g.][and references
  therein]{Kormendy13}. Simulations that fully evolve the bar potential do
show similar results as soon as they reach a quasi-steady state, i.e. after
the bar growth transient\footnote{Although promising, the coarse resolution of
  cosmological runs makes hard to fully resolve the nuclear region where the
  ILR is expected to occur.}. If the gas inflows accumulates enough mass at
$\sim R_{\rm ILR}$ the central region can dynamically decouple, possibly
forming nested non-axisymmetric structures (e.g. nuclear bars). These
structures can eventually bring the gas closer and closer to the galactic
centre in a cascade-like fashion \citep{Shlosman89}.

In this paper we propose a new set of fully evolving isolated galaxies
runs. We start with an unbarred galactic disc composed of stars and gas,
embedded in an evolving dark matter halo. We check the dependences of the gas
dynamics on different numerical implementations, varying the magnitude of an
artificial viscosity (if present) and the numerical resolution (see
section~\ref{suite} for a full description of the different runs). We run our
simulations
without implementing any gas radiative cooling, star formation and stellar
feedback prescriptions (usually referred to as sub-grid physics), in order to
perform a clean test of the basic numerical method used, and to highlight the
physical and purely dynamical effect of the forming substructures
(stellar spirals and bar) onto the gas.

As will be detailed we find that the flux of gas reaching the most central
regions of the galaxy peaks during the bar formation phase, and not when the
bar is fully established, independently of the exact numerical implementation.
We describe in details the set-up of our initial conditions and the features
of the simulation suite in Section \ref{section2}.  We present our main
findings in Section \ref{section3}, and we finally discuss them and derive our
conclusions in Section \ref{section4}, highlighting the relevance of our work
for the interpretation of observations and also commenting on the possible
shortcomings.


\section{Numerical methods}\label{section2}

\subsection{Initial conditions}

We simulate the isolated disc galaxy model Lmd2c12 described by
\citet{Mayer04}, in order to reproduce an initially bulgeless, bar-unstable
disc galaxy.  The galaxy model is made of three different components: a dark
matter halo, a stellar and a gaseous disc.

The dark matter halo follows the \citet[NFW][]{Navarro96, Navarro97} density
profile:
\begin{equation}
\rho_{\rm h} (r)= \frac{\rho_{\rm crit}~\delta_{\rm c}}{(r/r_{\rm s}) (1+r/r_{\rm
s})^2},
\end{equation}
where $r_{\rm s}$ is the scale radius of the halo, $\rho_{\rm crit}$ is the
critical density of the Universe today\footnote{We assume $H_{0} =
  71$~km~s$^{-1}$~Mpc$^{-1}$, compatible with the \emph{Wilkinson Microwave
    Anisotropy Probe} 7/9 years cosmology \citep{Komatsu11,Hinshaw13}}, and:
\begin{equation}
\delta_{\rm c}=\frac{200}{3} \frac{c^3}{\log(1+c)-c/(1+c)},
\end{equation}
depends only on the concentration parameter $c \equiv r_{200} / r_{\rm s}$.
$r_{200}$ is the radius that encompasses an average density $\langle \rho
\rangle = 200~\rho_{\rm c}$ and defines the outer radius of the dark matter
halo.  The mass of the halo is therefore $M_{200} = 200~\rho_{\rm c}~(4
\pi/3)~r_{\rm vir}^3$.  We adopt $c=12$ and a scale velocity $v_{200} =
\sqrt{G M_{200} / r_{200}} = 75$~km~s$^{-1}$, which corresponds\footnote{These
  numbers are slightly different from those reported by \citet{Mayer04}
  because of the different cosmology assumed. However, this does not affect
  the evolution of the galaxy model.}  to $M_{200} = 1.4 \times
10^{11}$~M$_{\sun}$, $r_{200} = 110$~kpc and $r_{\rm s} = 9.2$~kpc.

Both stellar and gaseous discs are modeled as a radial exponential disc with a
vertical structure modelled by isothermal sheets \citep{Hernquist93}:
\begin{equation} 
\rho_{\star}(R, z) = \frac{M_{\star}}{4 \pi R_{\star}^2 z_{\star}}
\exp(-R/R_{\star})~\cosh^{-2} \left( \frac{z}{z_{\star}} \right),
\end{equation}
where $R_{\star}=3$~kpc is the radial scale length and $z_{\star}=0.3$ kpc is
the vertical scale height.  The stellar disc has a total stellar mass
$M_{\star}= 1.4 \times 10^{10}$~M$_{\sun}$ and extends up to $10 R_{\star}$.
The gas component has a mass $M_{\rm gas} = 0.05 M_{\star} = 7 \times
10^8$~M$_{\sun}$ and its density profile is characterized by the same
parameters $R_{\star}$ and $z_{\star}$.  The gas has a uniform temperature
$T_0 = 10000$~K and we assume that it is composed of a mixture of ionized
hydrogen and helium with a mean molecular weight $\mu \simeq 0.59$.  All the
parameters are chosen in agreement with the galaxy-halo scalings predicted by
the $\Lambda$-CDM model \citep[e.g.][]{Mo98}.

We build the initial conditions using the code GINCO\footnote{GINCO (Galaxy
  INitial COnditions, \url{http://www.ics.uzh.ch/\~fiacconi/software.html}) was
written by Davide Fiacconi.}.  GINCO initializes
quasi-equilibrium galaxy models following \citet{Hernquist93} and
\citet{Springel05a}.  The models can be made of four arbitrary components: a
NFW dark matter halo, an exponential stellar and gaseous disc, and a spherical
bulge with the profile proposed by \citet{Hernquist90}. The polar/spherical
coordinates of the particles that belong to each component are randomly
sampled using the density profiles as probability distribution
functions. Then, polar/spherical angles are randomly drawn from isotropic
distributions and they are used to determine the Cartesian coordinates of the
particle positions.

The velocities are sampled from local Gaussian approximations of the true
distribution function \citep{Hernquist93}. The position-dependent parameters
of the Gaussians are computed solving the steady-state Jeans equations with
some closure assumptions on the velocity dispersion tensor \citep[see
  e.g.][]{Binney08}.  For spherical components (i. e. the dark matter
halo, since the simulated system is bulgeless), we assume that the velocity
dipersion tensor is isotropic (i.e. of
the form $\sigma^2(r)\,\mathbb{I}$, where $\mathbb{I}$ is the identity
matrix), with the 1D velocity dispersion given by:
\begin{equation}
 \sigma^2(r) = \frac{1}{\rho(r)} \int_{r}^{+\infty} \rho(x) \frac{{\rm d} \Phi_{\rm
tot}}{{\rm d}r}(x)~{\rm d}x,
\end{equation}
where $\rho(r)$ is the density profile of the considered component and
$\Phi_{\rm tot}$ is the total gravitational potential. The spherical
components have no net rotation.  The potential of the halo is an
analytic function; instead, the potential of the disc is computed as a
first-order vertical perturbation of the potential of a razor-thin exponential
disc, namely $\Phi_{\rm d}(R,z) \simeq \Phi_{0}(R) + \Phi_{1}(R,z)$.  The
razor-thin disc has the potential:
\begin{equation}
 \Phi_0(R) = -\frac{G M_{\star}}{R_{\star}}~y~\left[ I_{0}(y) K_{1}(y) - I_{1}(y)
K_{0}(y) \right],
\end{equation}
where $y = R/(2 R_{\star})$ and $I_i$ and $K_i$ are modified Bessel functions;
the first-order vertical perturbation is \citep[e.g.][]{Binney08}:
\begin{eqnarray}
 \Phi_1(R,z) & \equiv & 4 \pi G \int_{0}^{z} {\rm d}z' \int_{0}^{z'} {\rm d}z''
\rho_{\star}(R, z'') \nonumber \\
& = & 4 \pi G~\rho(R,0)~z_{\star}^2~\log \left[ \cosh \left( \frac{z}{z_{\star}}
\right) \right].
\end{eqnarray}
We use this strategy to maintain all the evaluations of the potentials and of
their derivatives analytic; this makes the code faster and reduces the
required memory.  Once we compute $\sigma^2(r)$, we can sample the magnitude
of the velocity of each particle in a spherical component from a Maxwellian
distribution with variance $\sigma^2(r)$.  Finally, we randomly draw the
spherical angles $(\theta, \phi)$ as above to ensure isotropy and we assign
the Cartesian components of the velocity.

\begin{figure}
\includegraphics[width=8.5cm]{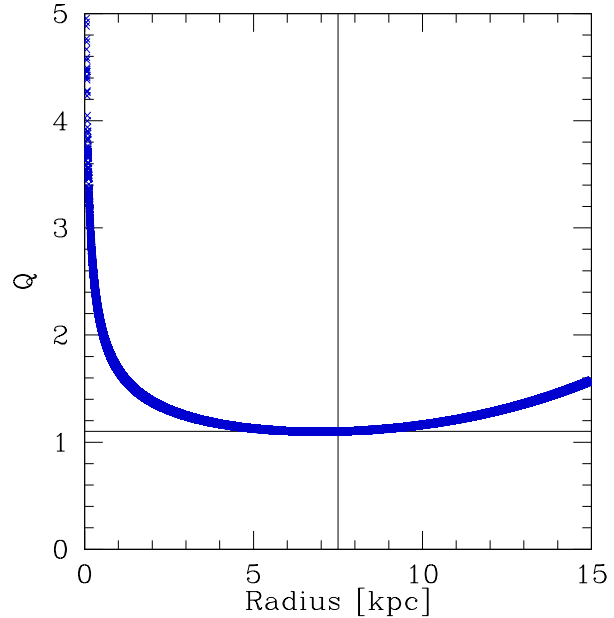}
\caption{Radial profile for the Toomre parameter of the initial stellar disc.}
\label{fig:figT}
\end{figure}

Both the stellar and the gaseous disc velocity structure is characterized by a
velocity dispersion tensor of the form ${\rm diag}(\sigma_r^2, \sigma_\phi^2,
\sigma_z^2)$.  The vertical velocity dispersion is
\citep{Hernquist93,Springel05a}:
\begin{eqnarray}
 \sigma_z^2(R,z) & = & \frac{1}{\rho_{\star}(R,z)} \int_{z}^{+\infty}
\rho_{\star}(R,z') \frac{\partial \Phi_{\rm tot}}{\partial z}(R,z')~{\rm d}z'
\nonumber \\
 & \approx & \frac{G M_{\star} z_{\star}}{2 R_{\star}^2} \exp
\left(-\frac{R}{R_{\star}} \right),
\end{eqnarray}
where the last approximation holds when the disc is geometrically thin
and the vertical gradient of the potential around $z \simeq 0$ is
dominated by the disc. The radial component is chosen to be
  $\sigma_r^2 \propto \sigma_z^2$, with the normalization enforcing
  a minimum value of the Toomre parameter $\mathcal{Q} \simeq 1.1$ at
  $r \simeq 2.5 R_{\star}$ \citep{Mayer04}. The whole profile of
  $\mathcal{Q}$ for our initial conditions is shown in
  Fig.~\ref{fig:figT}.  The azimuthal component is set using the
  epicyclic approximation, $\sigma_\phi^2 = \sigma_r^2~\kappa^2 / (4
  \Omega^2)$. Unlike the dark matter halo, the disc has a net
rotation, i.e. an average azimuthal velocity $\langle v_{\phi}
\rangle$ given by \citep{Hernquist93,Springel05a}:
\begin{equation}
 \langle v_\phi \rangle^2 = V_{\rm c}^2 + \sigma^2_r \left( 1 - \frac{\kappa^2}{4
\Omega^2} - \frac{2R}{R_{\star}} \right),
\end{equation}
where $V_{\rm c}$ is the circular velocity in $\Phi_{\rm tot}$.
Finally, we sample the $(v_r, v_\phi, v_z)$ components of the velocity of each
disc (both star and gas) particles from gaussian distributions with mean
$(0,\langle v_\phi \rangle,0)$ and standard deviations $(\sigma_r,
\sigma_\phi, \sigma_z)$, respectively, and we finally transform then into the
Cartesian components.

\begin{figure*}
\begin{minipage}{0.48\textwidth}
  \centering
  \includegraphics[width=8cm]{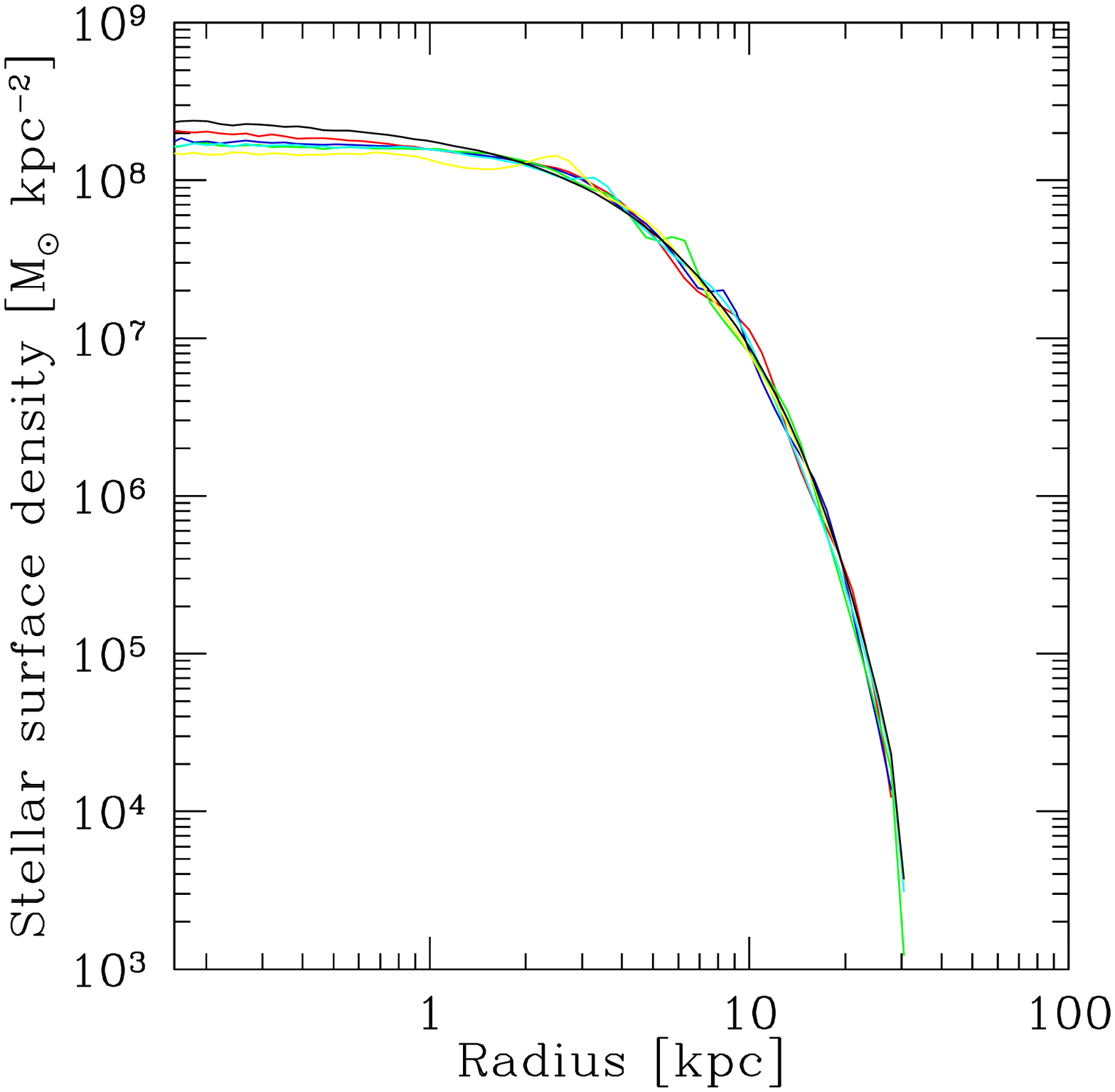}
\end{minipage}
\hfill
\begin{minipage}{0.48\textwidth}
  \centering
  \includegraphics[width=8cm]{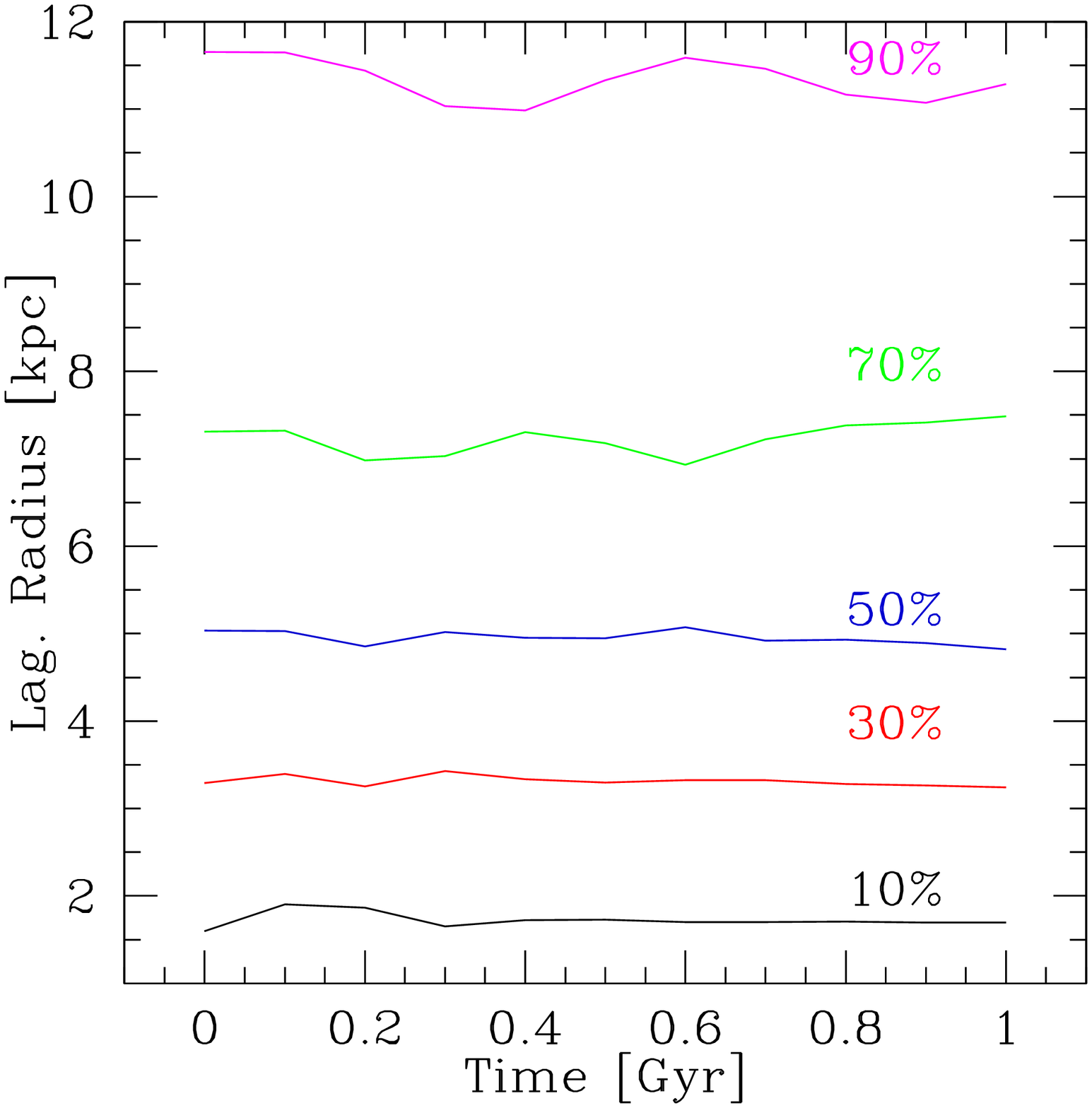}
\end{minipage}
\caption{Left panel: surface density profile of the stellar component in the first
Gyr. Black, yellow, cyan, green, blue and red curves correspond to
   $t=$0, 0.2, 0.4, 0.6, 0.8 and 1 Gyr, respectively. 
   Right panel: lagrangian radius at different stellar mass fraction in the first
Gyr. Black, red, blue, green and magenta lines represent $10\%$, $30\%$, $50\%$,
$70\%$, $90\%$ of stellar mass, respectively.}
\label{fig:figS}
\end{figure*}

We checked the stability of our initial conditions studying the
  evolution of the stellar surface density profile as a function of
  time (left panel of Fig.~\ref{fig:figS}) during the first Gyr,
  i.e. before the development of strong non-axisymmetric perturbation
  (see below). After a short transient phase due to the non-exact
  equilibrium of the initial conditions (highlighted be the yellow
  line in figure) the system re-adjust on a profile similar to the
  initial one, with the surface density at $t=1$ Gyr (red line)
  differing by $20 \%$ at most with respect to the initial conditions
  (within the disc scalelength).  Similar conclusions about the
  stability of the stellar disk can be drawn from the evolution of its
  Lagrangian radii (right panel of Fig.~\ref{fig:figS}).

\subsection{Simulation suite}\label{suite}

We run a suite of numerical simulations of the reference model described in
the previous Section in order to explore the effects of resolution, numerical
implementation and parametrization of the artificial viscosity (when present).
In Table \ref{tab:tab1} we summarize the sample of 3D runs presented in this
work.  We build two realizations of our initial conditions at two resolutions:
\begin{enumerate}
\item low resolution (LR): the halo is sampled with $10^6$ particles with mass
  $m_{\rm h} = 1.4 \times 10^5$~M$_{\odot}$, while the stellar and gaseous
  discs are sample with $9.5 \times 10^5$ and $5 \times 10^4$ particles,
  respectively, with mass $m_{\star} = m_{\rm gas}\simeq 1.5 \times
  10^4$~M$_{\odot}$.  The gravitational softenings (setting the spatial/force
  resolution of the gravitational interaction) for dark matter and baryonic
  particles (equal for stars and gas particles) are $65$~pc and $20$~pc,
  respectively;
\item high resolution (HR): the halo is sampled with $8 \times 10^6$ particles with
  mass $m_{\rm h} = 1.6 \times 10^4$~M$_{\odot}$, while the stellar and
  gaseous discs are sample with $7.6 \times 10^6$ and $4 \times 10^5$
  particles, respectively, with mass $m_{\star} = m_{\rm gas} \simeq 1.7 \times
  10^3$~M$_{\odot}$.  The gravitational softenings for dark matter and
  baryonic particles are $30$~pc and $7$~pc, respectively.
\end{enumerate}
We ensure that the particles in the disc (star and gas) have all the same
mass, preventing any spurious relaxation/mass segregation.  All the
simulations assume an isothermal equation of state to simply model an
effective atomic radiative cooling keeping the ISM in the disc plane at an
almost constant temperature $\lesssim 10^4$~K. Metal line and molecular
cooling would reduce the gas temperature further, allowing for dense clumps to
form and to trigger star-formation. Feedback from stars would then re-heat the
gas, resulting in the formation of a multi-phase medium\citep[e.g.][]{Wada01,
  WadaNorman01}. Because of the lack of cooling and star-formation physics, we
keep an high temperature to prevent the sudden fragmentation of the gaseous
disc.

We test the robustness of our results against two different implementations of
the hydrodynamics.  Most of the simulations are performed with the
Tree/Smoothed Particle Hydrodynamics (SPH) code {\sc gadget2}
\citep{Springel05b}, which uses an oct-tree structure to speed up the gravity
calculations \citep{Barnes86} and threats the hydrodynamics with the
density-entropy SPH proposed by \citet{Springel02}.  The SPH formalism
requires the introduction of an artificial viscosity in order to capture
shocks correctly \citep[e.g.][]{Monaghan92,Balsara95,Monaghan97}. Therefore,
we explore the effect of different choices of the value of the artificial
viscosity parameter $\alpha$\footnote{The $\beta$ parameter in the
  Monaghan-Balsara formulation is equal to 2 $\alpha$ in all our
  runs.}. Finally, we also compare the results from SPH simulations with a run
that uses the newly developed code {\sc gizmo} \citep{Hopkins14}. {\sc gizmo}
is a meshfree code that captures advantages from both SPH and grid codes: it
preserves the Lagrangian structure of SPH codes, but at the same time solved
directly the Euler equations among different regions of the computational
domain without requiring the implementation of any artificial viscosity. We
used it in its finite-mass variant, in which there is not mass flux among the
regions belonging to different particles, keeping the mass of each gas
particle fixed.


\begin{table}
\caption{Summary of simulations characteristics. 
Columns: (1) name of the simulation, (2-3) resolution, (4) code used, (5) artificial
viscosity $\alpha$.}
\label{tab:tab1}
\begin{tabular}{lcccc}
\hline
Name & Barion particle &  DM particle    & Code & $\alpha$ \\
     &    softening (pc)    &  softening (pc)     &      &          \\
\hline
LR & 20 & 65 &{\sc gadget2} & 0.8 \\
LRV16 & 20 & 65 & {\sc gadget2} & 1.6 \\
LRV04 & 20 & 65 & {\sc gadget2} & 0.4 \\
LRGiz & 20 & 65 & {\sc gizmo} & - \\
HR & 7  & 30 & {\sc gadget2} & 0.8 \\
\hline
\end{tabular}
\end{table}


\section{Results}\label{section3}

\subsection{Low resolution simulations}

\begin{figure*}
        \centering
                \hspace{-0.5cm}
                \includegraphics[width=19.cm,clip=true,trim=0 12.8cm 0 0]{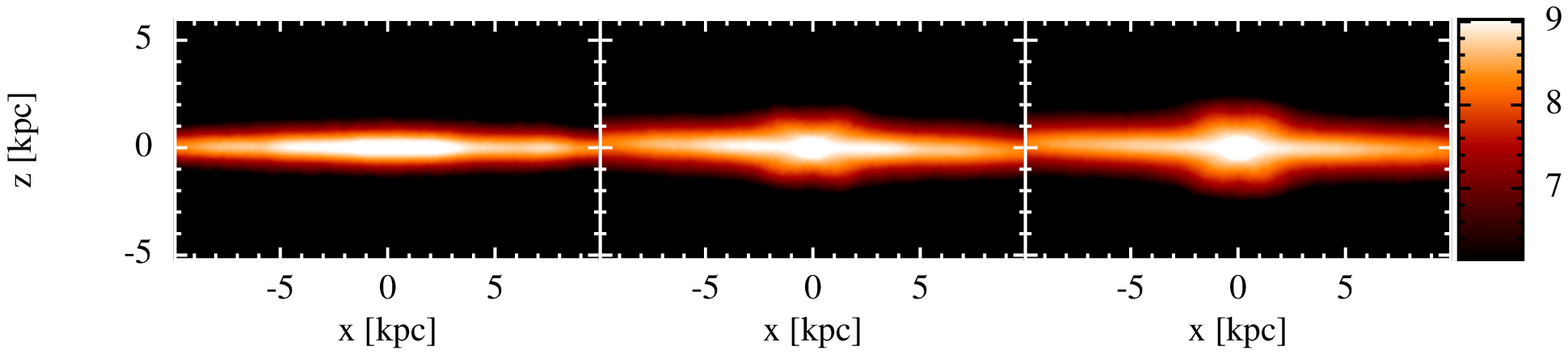}
                \includegraphics[width=19.cm,clip=true,trim=0 10.6cm 0 0.1cm]{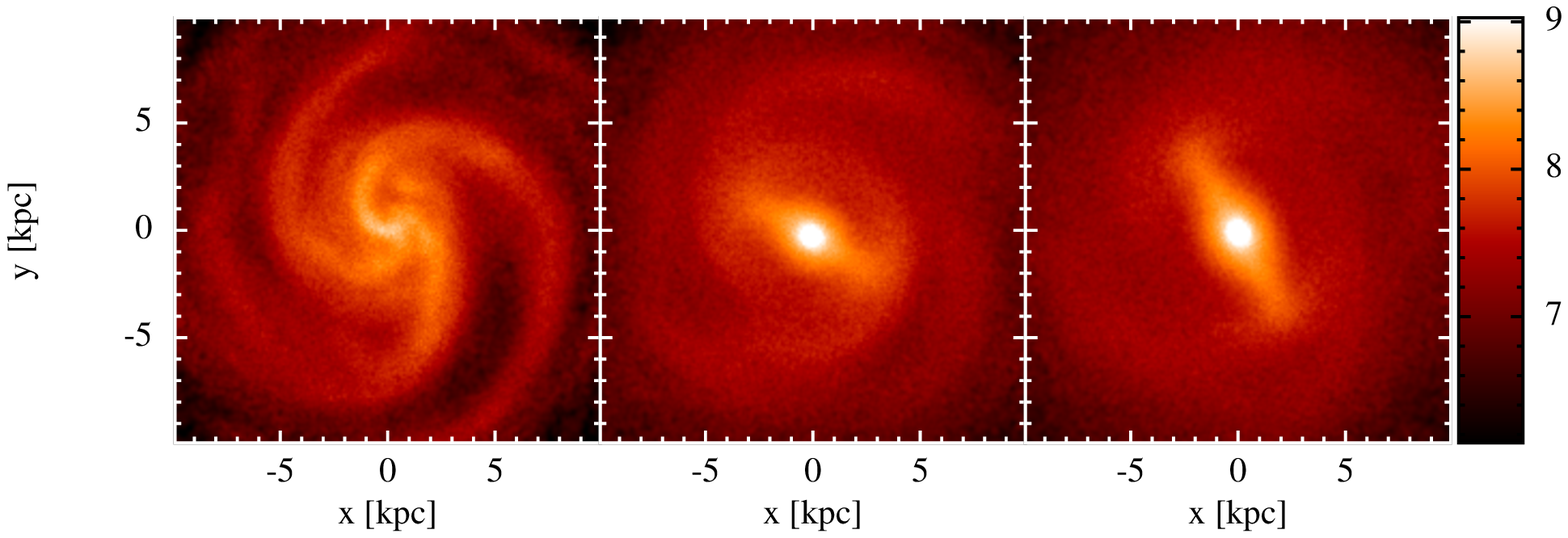}
                \includegraphics[width=19.cm,clip=true,trim=0 8cm 0 0.1cm]{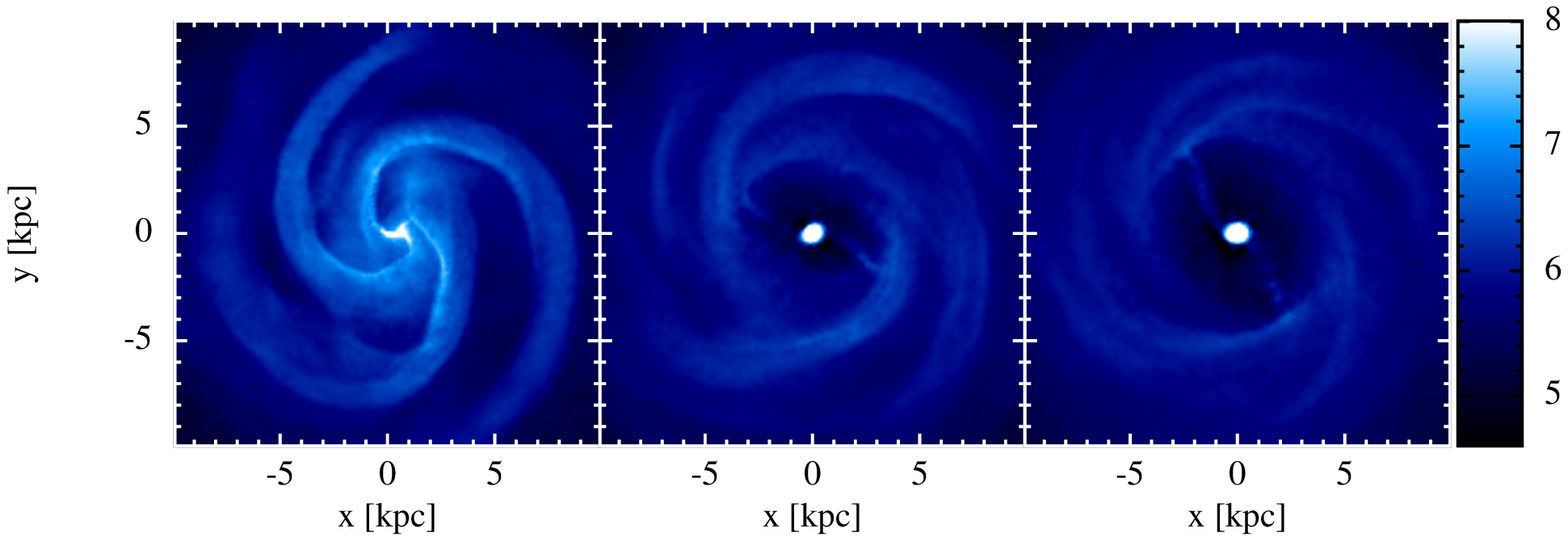}

\caption{Upper (middle) panels: edge-on (face-on) views of the stellar disc at
  $t=$1, 4 and 7 Gyr (left, central and right panel, respectively). The colour
  gradient maps the stellar surface density (in units of M$_{\odot}$
  kpc$^{-2}$) on a logarithmic scale. Bottom panels, same as the middle panel
  for the gas surface density.}
\label{fig:fig1}
\end{figure*}

\begin{figure}
\includegraphics[width=8.5cm]{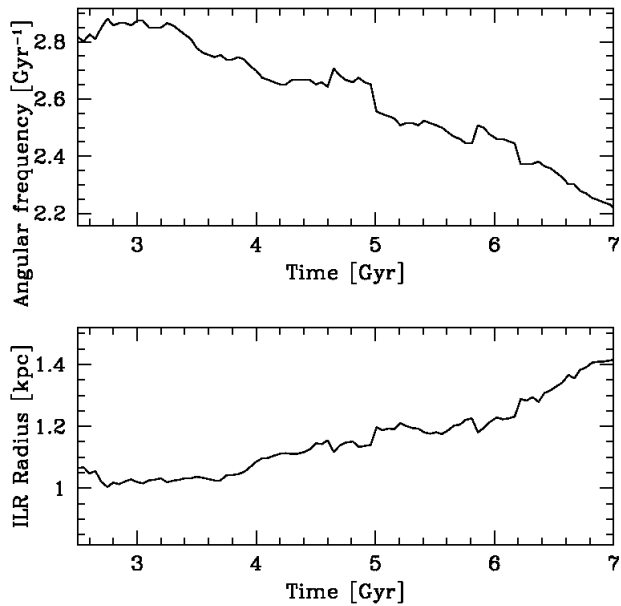}
\caption{Angular frequency of the bar (upper panel) and radius corresponding to the
inner Lindblad resonance from t=2.5 to t=7 Gyr (lower panel).}
\label{fig:fig2}
\end{figure}

Figure~\ref{fig:fig1} shows the distribution of star and gas observed in the 
LR run at three different times, $t=$1, 4 and 7 Gyr in the left, central and
right panels respectively. The stellar surface density is shown in the upper
and middle panels (edge-on and face-on views, respectively), while the face-on
view of the gas surface density is shown in the lower panels.
During the first 2 Gyr the bar-unstable system evolves from a axisymmetric
configuration to a barred disc, passing through the formation of transient
multi-arm spirals. In particular, a three arm spiral structure is observable
in the stellar density distribution at $t=$1 Gyr in the left-middle panel of
figure~\ref{fig:fig1}. From 2 Gyr on the disc shows a clear bar structure
(with a size of about 8 kpc) in its central region. From the bar-formation
time ($t\approx 2$ Gyr) on, the bar tends to slow-down, as shown in the upper
panel of figure~\ref{fig:fig2}. At $t \lsim 3$ Gyr the bar makes almost 2.8
full precessions per Gyr, while the frequency decreases down to $\lsim 2.3$
precessions per Gyr at $t\approx 7$ Gyr. The bar slow-down, already extensively
discussed in literature \citep[e.g.][]{Sellwood81, Combes81, Halle15}, results
in a $R_{\rm ILR}$ growing in time, from $\sim 1$ kpc up to $\sim 1.4$ kpc
at the end of the run, as observable in the lower panel of figure~\ref{fig:fig2}.
The bar forms thin, and buckles in its centre as the
time goes by, as observable in the edge on view of the stellar disc at $t=$4
and 7 Gyr. At the end of the simulation a boxy-peanut bulge like structure is
observable within the central few kpc of the disc.

The dynamics of the subdominant gas component is dominated by the underlying
stellar dynamics. During the first 2 Gyr the gas distribution resembles the
stellar one, with clear spiral arms (almost co-spatial with the stellar ones)
observable (see the example in the left lower panel of figure~\ref{fig:fig1}
at $t=$1 Gyr). After the formation of the stellar bar, the gas within the bar
corotational radius ($R_{\rm C} \approx$ 4-5 kpc depending on the age of the bar, as
will be discussed below) is driven toward the galaxy centre, and forms a dense
knot of gas clearly observable in the central and right panels in the bottom
row of figure~\ref{fig:fig1}. The torquing effect of the spiral arms before
and the stellar bar afterwards sweeps the almost totality of the gas between
$R_{\rm C}$ and the central dense knot. A small amount of low dense gas is still
observable in this "dead region", in particular in the form of two inflowing
streams connecting the outer galactic disc with the central dense knot, often
observed in simulations as well as in real galaxies \citep[e.g.][]{Regan99}.

\begin{figure*}
\begin{minipage}{0.48\textwidth}
  \centering
  \includegraphics[width=8cm]{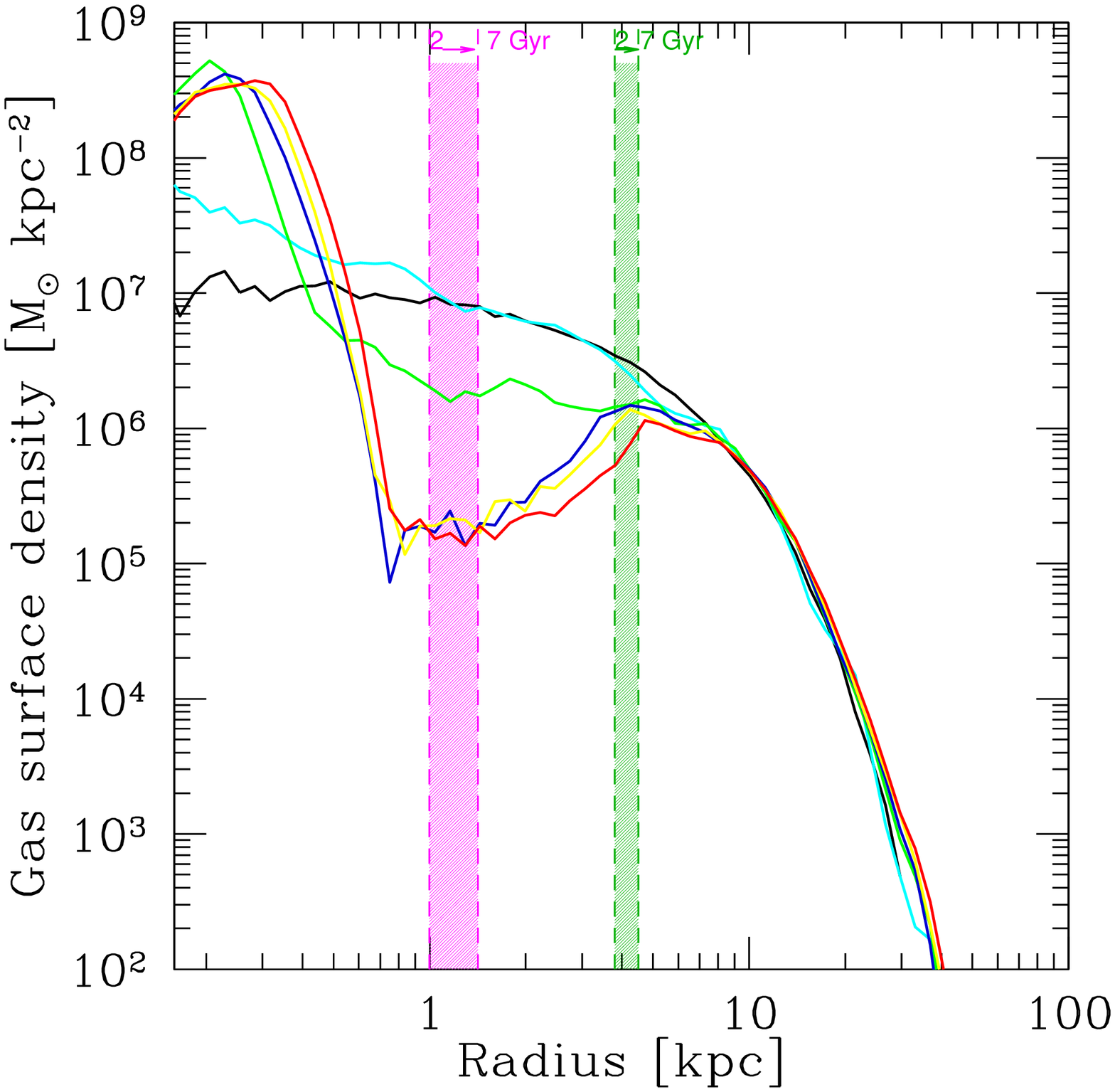}
\end{minipage}
\hfill
\begin{minipage}{0.48\textwidth}
  \centering
  \includegraphics[width=8cm]{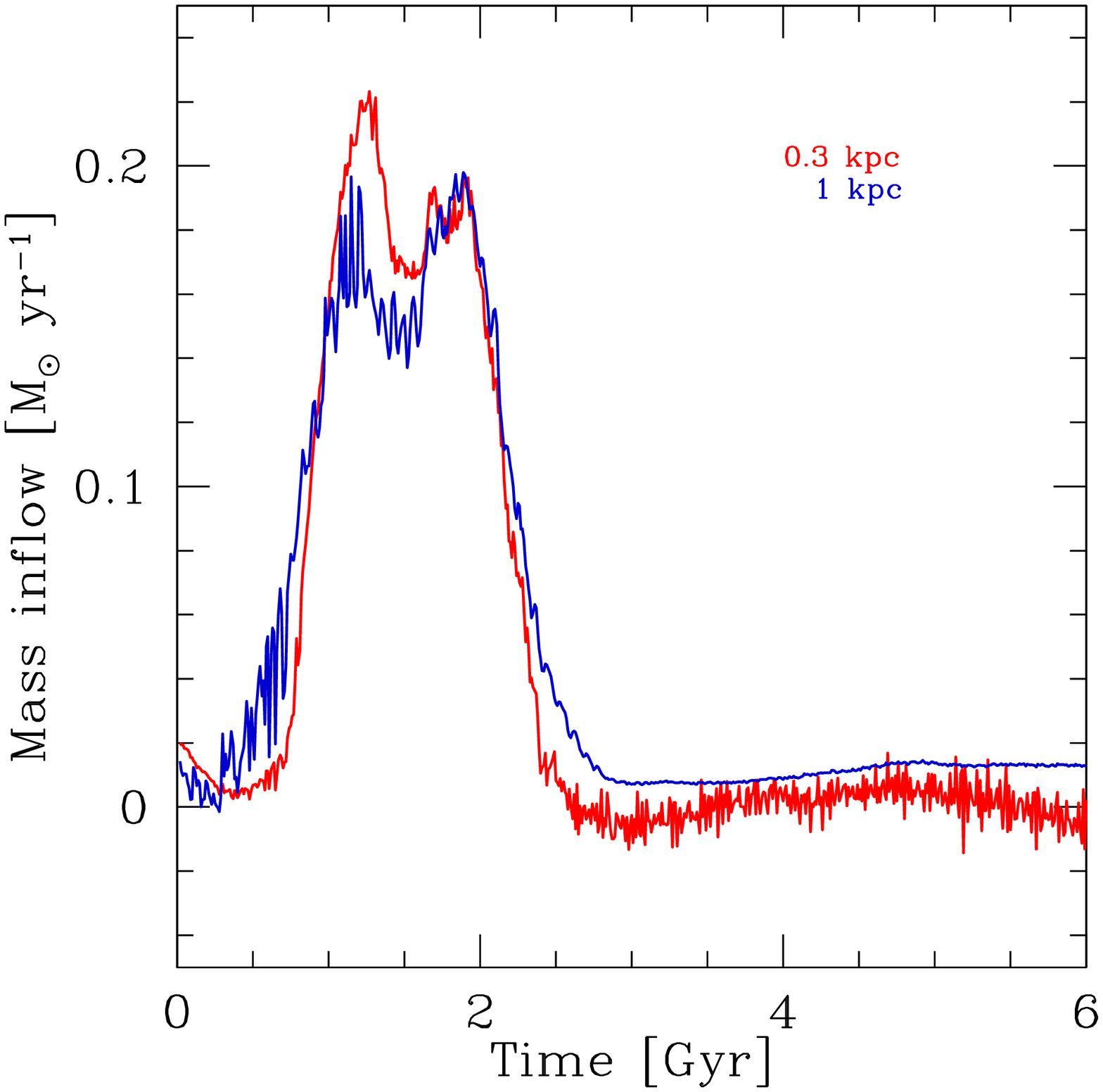}
\end{minipage}
\caption{Left panel: surface density profile of the gaseous component. Solid black,
cyan, green, blue,
   yellow and red curves correspond to
   $t=$0,1,2,3,5 and 7 Gyr, respectively. The shaded magenta and green areas mark
the region span
   by $R_{\rm ILR}$ and $R_{\rm C}$, respectively.
   Right panel: gaseous mass inflow as a function of time. Red and
   blue lines represent the accretion rate computed at 0.3 and 1 kpc from the
   center respectively.}
\label{fig:fig3}
\end{figure*}

The left panel of figure~\ref{fig:fig3} quantifies the effect that the bar
formation process has onto the gas. The surface density of the gas in the dead
zone decreases by up to $\sim 1.5$ orders of magnitude at $t\gsim 3$ Gyr
(blue, yellow and red lines) with respect to the initial conditions (black
line).  The shaded areas in figure trace the evolution of $R_{\rm C}$ (green) and
its outermost inner Lindblad resonance radius ($R_{\rm ILR}$ pink), from when
a clear bar structure is observable and its angular frequency is measurable
($t\approx 2$ Gyr) to the end of the simulation. The gas within $R_{\rm C}$ is
dragged toward scales of the order of $R_{\rm ILR}$, fueling the formation of
the central knot of gas on sub-kpc scales \citep[in agreement with a wealth of
  previous studies, e.g.][]{Sanders76, Shlosman89, Athanassoula92,
  Berentzen98, Regan04, Kim12, Cole14}, where the surface density increases by
up to almost 2 orders of magnitude.

A clear although less obvious result of the LR run consists in the
efficiency of the ``dead zone" formation. Most of the inflow from
$R<R_{\rm C}$ to $R\lsim R_{\rm ILR}$ happens during the first 2 Gyr,
as observable comparing the cyan ($t=$1 Gyr), green ($t=$2 Gyr) and
blue ($t=$3 Gyr) lines with the initial conditions and the end result
of the simulation in the left panel of figure~\ref{fig:fig3}. The
fully formed bar does indeed play a role in further decreasing the gas
surface density on the dead zone, and most importantly, in preventing
new gas to refill the central regions by pushing the gas immediately
outside the CR toward the outer Lindblad resonance radius ($R_{\rm
  OLR}$).\footnote{Although harder to be noticed in a log-log plot,
  the gas surface density decreases in the $R_{\rm C}<R<R_{\rm OLR}$
  region, and the material accumulates just outside $R_{\rm OLR}$.}
However, it is instead the formation of the bar which is efficient in
driving sustantial gas inflow. The fundamental importance of the
torques acting on the gas during the build-up of the bar, before this
has been fully developed, is highlighted in the right panel of
figure~\ref{fig:fig3}, in which we show the gas accretion rate
$\dot{M}$ within $R_{\rm C}$ as a function of time. In particular, the
red and blue lines refer to $\dot{M}$ through surfaces at 0.3 and 1
kpc from the centre, respectively. At both scales $\dot{M}$ shows a
first prominent peak at $t \approx 1$ Gyr, well before the formation
of any significant bar-like structure. A second peak of similar
magnitude is observable at $t\approx 2$ Gyr, just after the bar has
formed, while the central fueling drops immediately
afterward. Although our simple simulation does not include any star
formation prescription, such omission has little impact on the
evolution of the gas from $R_{C}$ down to the nuclear knot, since the
majority of the inflow happens on a few (up to $\approx 10$ close to
$R_{\rm ILR}$) orbital timescales. The time evolution of the accretion
flows through the two surfaces is quite similar at all times.

\subsection{Viscosity test}

As recently reviewed by \cite{Sellwood14}, the gas angular momentum transport
in SPH simulations could be at least partially affected by the artificial
viscosity used. Differently from grid based codes, in which a numerical
viscosity is intrinsically related with the discretization of the space
domain, in SPH codes the numerical viscosity is explicitly taken into account
through a viscosity parameter $\alpha$. The shear and bulk viscosity in SPH
simulations scale linearly with $\alpha$ \citep[e.g.][]{Murray96, Lodato10}.

In this section we test the effect of the artificial viscosity through the
comparison of run LR with three different simulations. Two of these,
LRV04 and LRV16, are exact copies of the LR run, but for the value of
the $\alpha$ parameter, that is half and double of the $\alpha=0.8$ value used
in LR. The third simulation (LRGiz) has been run using the GIZMO code
\citep{Hopkins14}, that solves the evolution of the gas on an unstructured
grid and does not require any explicit artificial viscosity term.

\begin{figure*}
\includegraphics[width=16cm]{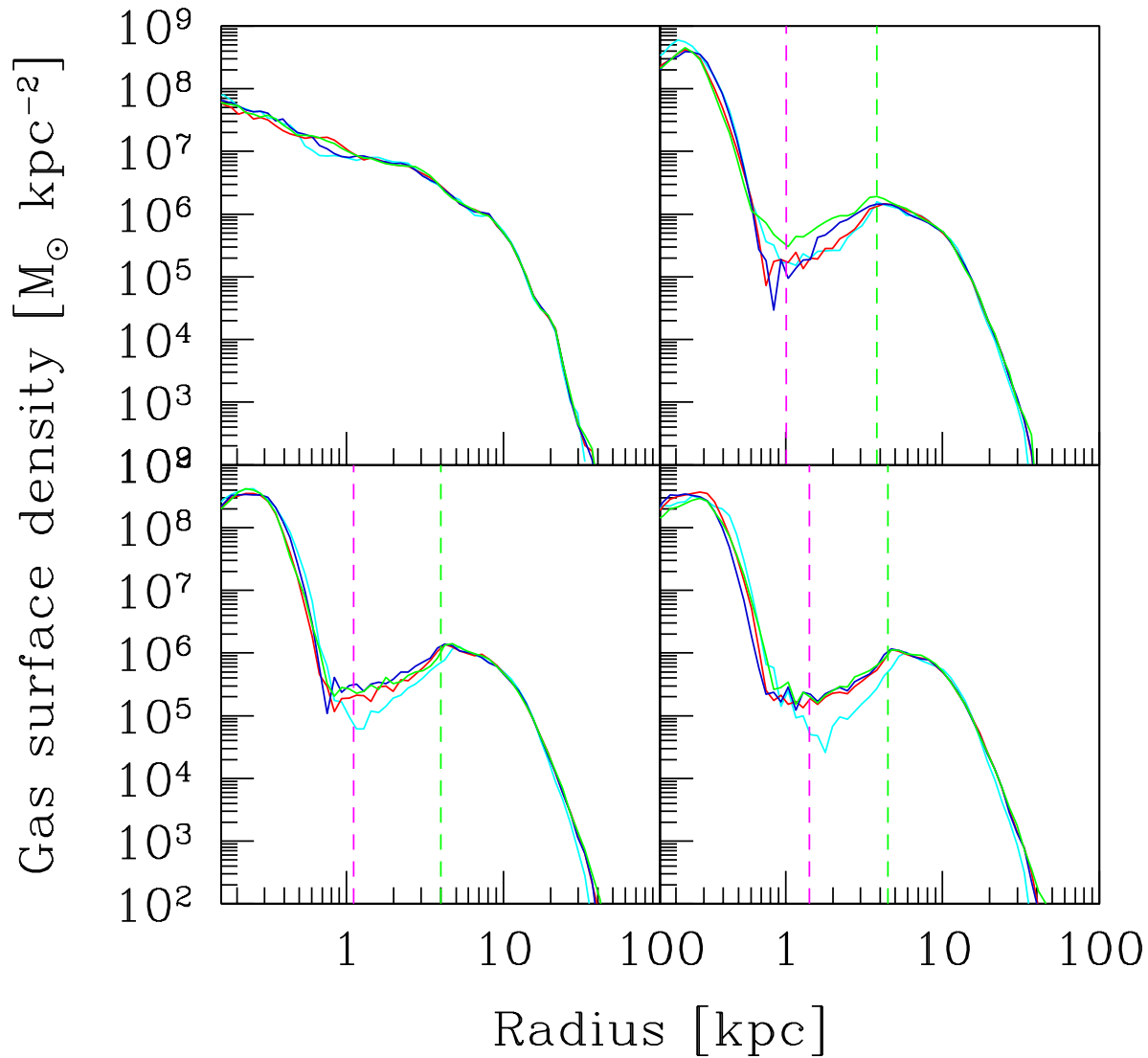}
\caption{Surface density profile for  low resolution
  isothermal simulations with different viscosity. Upper left, upper right,
  lower left and lower right panels refer to t=1 Gyr, 3 Gyrs, 5 Gyrs and 7
  Gyrs, respectively.  The solid red, green and blue curves are obtained from
  the gas particle distribution for simulation with $\alpha=0.8$ (run LR),
  $\alpha=1.6$ (run LRV16) and $\alpha=0.4$ (run LRV04). The cyan curve
  corresponds to the gas particle distribution for simulation using GIZMO
  (LRGiz). The dashed magenta and green lines mark the positions of $R_{\rm
    ILR}$ and $R_{\rm C}$ at the different times.}
\label{fig:fig4}
\end{figure*}

\begin{figure*}
\includegraphics[width=20cm]{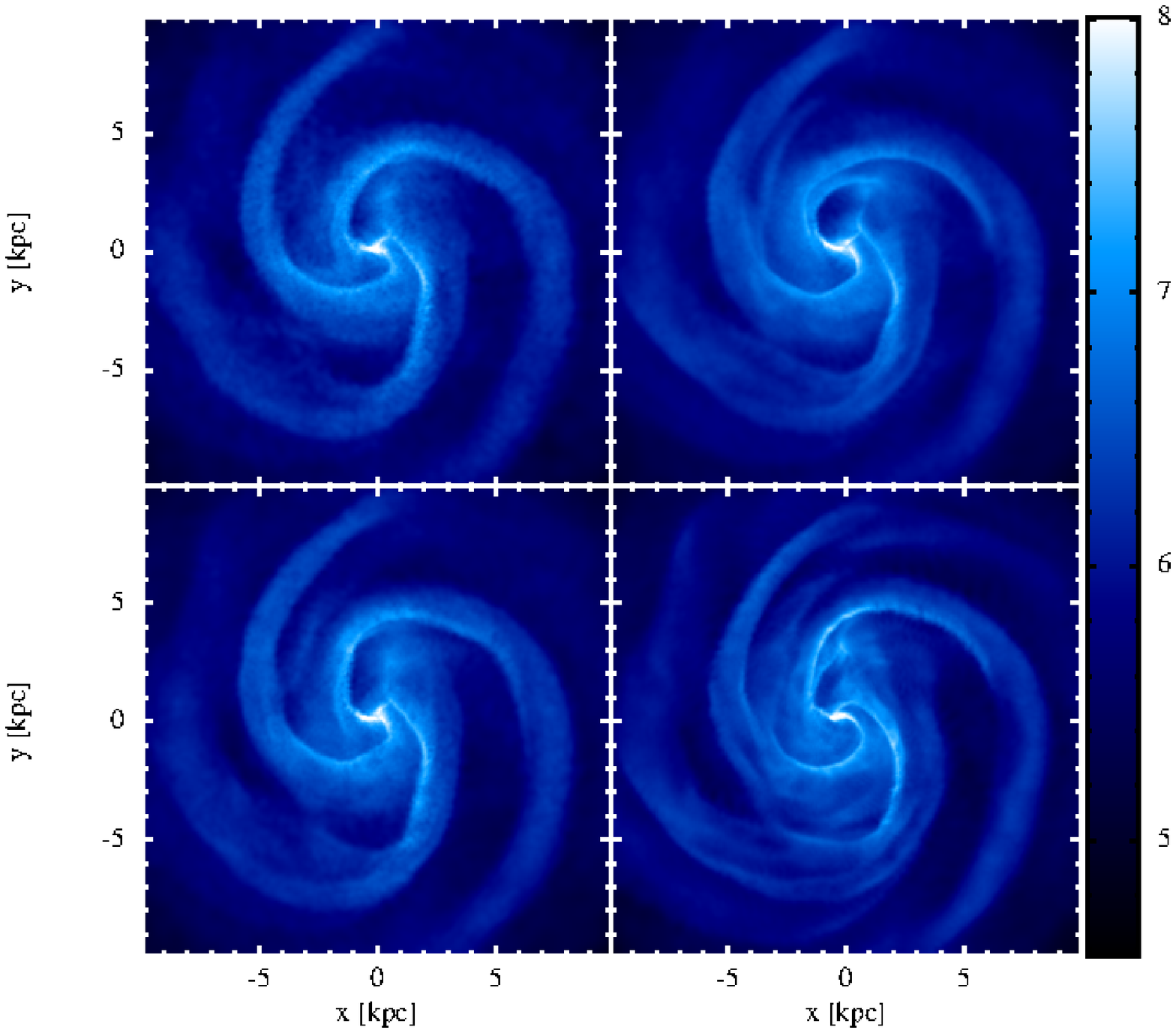}
\caption{Face-on views of gas surface density for  low resolution simulations with
different values of $\alpha$ at t=1 Gyr: $\alpha$=0.4 in the left upper panel,
$\alpha$=1.6 in the right upper panel, $\alpha$=0.8 in the left lower panel and  low
resolution simulation using GIZMO in the right lower panel. The logarithmic density
scale is in units of $\rm M_{\rm \odot}$ $\rm kpc^{-2}$.}
\label{fig:fig5}
\end{figure*}

The results of the test are shown in figure~\ref{fig:fig4} and
figure~\ref{fig:fig5}. Figure~\ref{fig:fig4} shows the comparison between the
surface density profiles of the four runs (LR in red, LRV16 in green,
LRV04 in blue and LRGiz in cyan) at four diffent times, $t=$1 Gyr (upper
left panel), 3 Gyr (upper right panel), 5 Gyr (lower left panel) and 7 Gyr
(lower right panel). Similarly, figure~\ref{fig:fig5} shows the face-on
projection of the gas surface density map for the four runs at $t=1$ Gyr, to
allow for a comparison of the non-axisymmetric structures forming. The
comparison between the three SPH runs shows that the exact value of the
viscosity parameter $\alpha$ plays a little role in the gas dynamics. The
removal of gas from the forming dead zone and the formation of a dense central
gas knot are completely dominated by the gravitational torques due to the
formation of non-axisymmetric structures. The LRGiz run shows some very minor
differences too. 



\subsection{High resolution simulations}

{\bf \subsubsection{Low vs high resolution run comparison}}

As a final test, we ran an increased resolution version of the LR
simulation (HR), as discussed in section~\ref{section2}. Because of
the higher spatial resolution and of the isothermal equation of state
implemented, the gas in the HR run forms extremely dense and compact
clouds in the galaxy nucleus, slowing down the simulation enormously
after the first episode of major gas inflow. For this reason we have
run the HR simulation only up to $t\approx 3$ Gyr, and we limit our
analysis to the response of the gas to the initial spiral and bar
formation.

\begin{figure*}
                        \begin{minipage}{.48\textwidth}
                        \centering
                                \includegraphics[width=11cm]{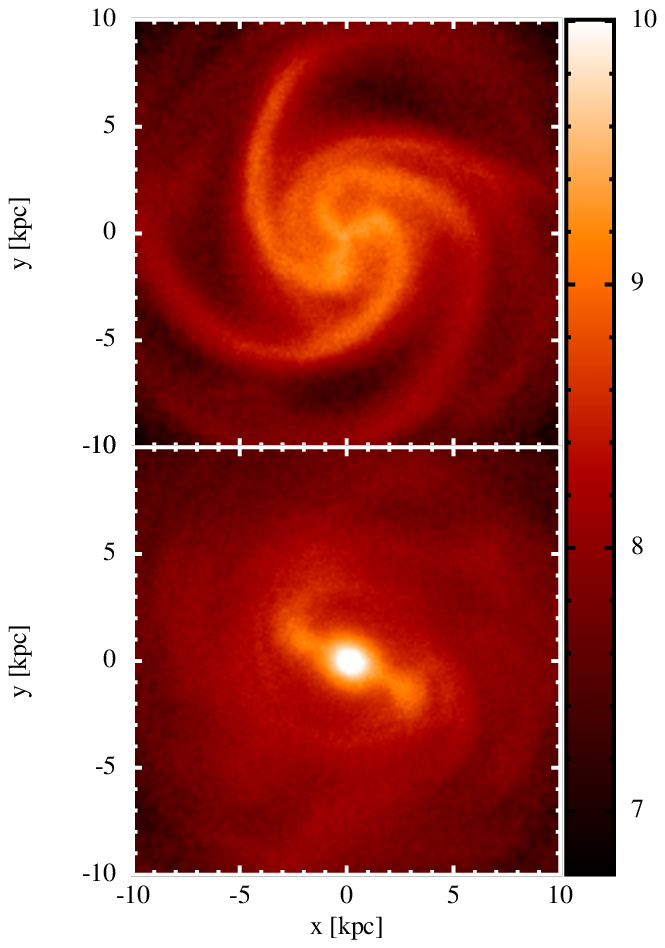}
                        \end{minipage}
                        \hfill
                        \begin{minipage}{.48\textwidth}
                        \centering
                                \includegraphics[width=9.2cm, clip=true,trim=1.3cm 0 0 0]{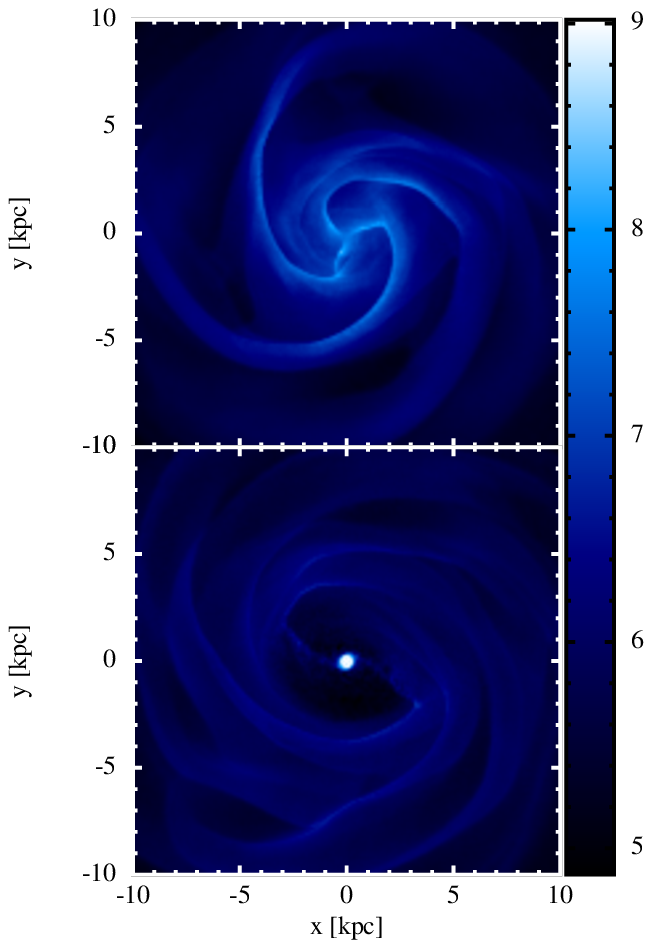}
                        \end{minipage}
 \caption{Right panels: logarithmic face-on views of the stellar surface density (in
unit of $\rm M_{\rm \odot}$ $\rm kpc^{-2}$) for the  high
   resolution simulation (HR) at $t=1$ Gyr (upper panel) and $t=2.5$ Gyr
   (lower panel). Left panels: same as right panels for the gas surface
   density.}
 \label{fig:fig6}
 \end{figure*}

Fig. \ref{fig:fig6} shows the face-on views of the stellar (left
panels) and gaseous (right panels) surface density. Similarly to the low
resolution simulation, during the first Gyr, the system develops stellar
spiral arms (upper left panel) which evolve in a stable bar like structure
(lower left panel) at about t=2.5 Gyrs. The gas follows a similar dynamics as
observed in the lower resolution runs, following the stellar spiral arms
during the first evolutionary phase and being driven toward the centre during
and after the bar formation. As in the other runs, in the central region of
the galaxy (within the bar extent) a dead zone forms, with very low density
gas present in between the outer disc and the nuclear gas knot.

\begin{figure*} 
\begin{minipage}{0.48\textwidth}
 \centering
 \includegraphics[width=8cm]{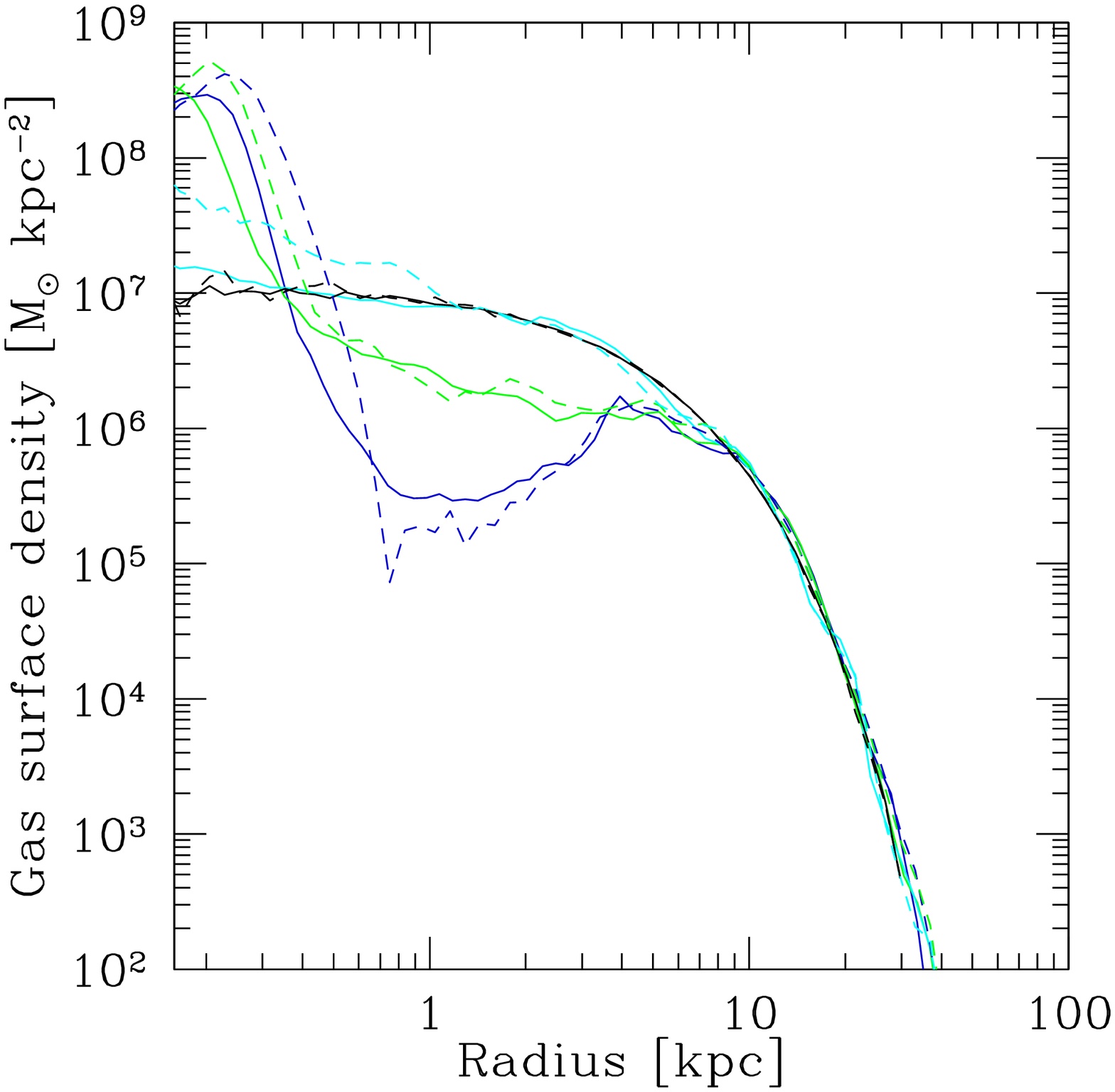}
\end{minipage}
\hfill
\begin{minipage}{0.48\textwidth}
 \centering
 \includegraphics[width=8cm]{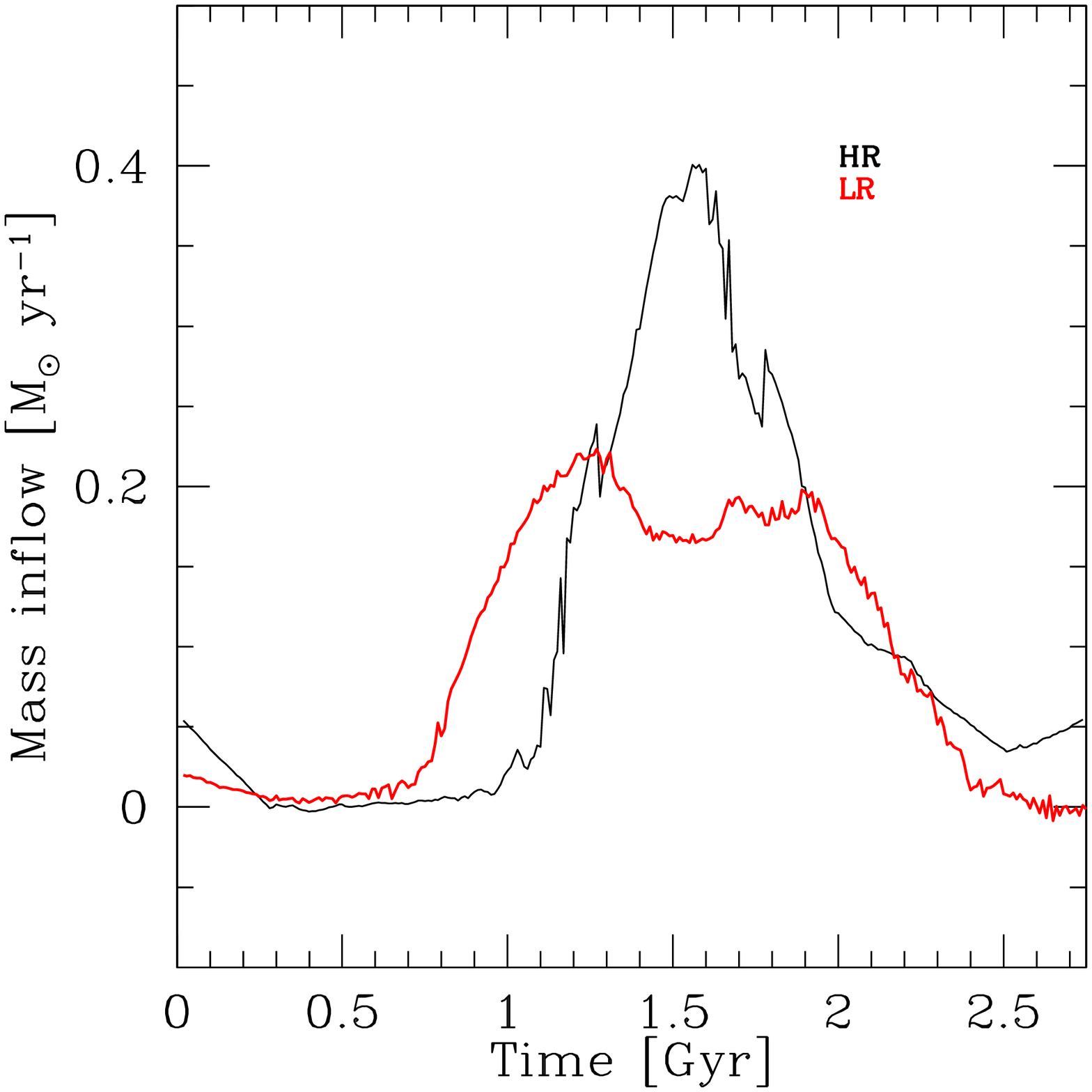}
\end{minipage}
\caption{Left panel: surface density profile of the gas. Solid lines refer to the
   highest resolution run HR. The results of the low resolution run LR
   are reported with dashed lines for comparison. Black, cyan, green and blue
   curves refer to $t=$0,1,2, and 3 Gyr, respectively.  Right panel: gaseous mass
   inflow as a function of time. Black and red lines represent the inflow rate
   computed at 0.3 kpc from the center in run HR and LR respectively.}
\label{fig:fig7}
\end{figure*}

Figure~\ref{fig:fig7} shows a comparison between HR and LR runs. The
left panel shows the gas surface density profiles in the two runs for four
different times. The biggest difference is observable at $t\approx 1$ Gyr,
when in the low resolution run the disc is already significantly perturbed,
while in the HR run the gas profile is still quite unperturbed. The gas
profile in the high resolution run is more similar to its low resolution
analogous at later times, but for a slightly more pronounced dead zone in the
HR run due to the better resolved profile of the stellar bar, that results
in a more effective action of the bar itself onto the gas.

The main difference observed in the profiles at $t\approx 1$ Gyr is due to the
later growth of non-axisymmetric perturbations (first in the form of spiral
arms, turning into a central bar) in the higher resolution simulation. This is
clearly observable in the accretion rate through the central 0.3 kpc (right
panel of figure~\ref{fig:fig7}). In run HR the peak of $\dot{M}$ occurs at
$t\approx 1.5$ Gyr, about 0.2-0.3 Gyr after the peak observed in the LR
run. Again, in the high resolution simulation the $\dot{M}$ peak has a larger
intensity (by almost a factor of 2) with respect to the low resolution case,
due to the more efficient cleaning of the dead zone during the bar formation
process. 

The later growth of non-axisymmetric structures in the higher resolution run
is probably due to the lower shot noise in the initial conditions: a higher
number of particle MonteCarlo sampling results in a lower statistical noise,
from which structures can grow \citep[see also the discussion
  in][]{Sellwood14}.  An extensive and time consuming study aiming
at numerical convergence is neither feasible (within the currently available
computational facilities) nor useful, as a simulation with a order of
magnitudes larger number of particles could result in a degree of symmetry
significantly larger than any real disc galaxy observed. The dependence of the
$\dot{M}$ peak and of the time at which spirals and bars form on the number of
particles used demonstrate that these should not be taken as physical
values. Only the gas response pattern is similar in all the runs analyzed,
independently of the viscosity prescription adopted, of the numerical
resolution achieved, and of the algorithm used to solve the gas dynamics.

\subsubsection{Dynamics of the nuclear inflow}

\begin{figure*}
        \centering
                \hspace{-1.6cm}
                \includegraphics[width=19.3cm,clip=true,trim=0.6cm 1.3cm 0 0]{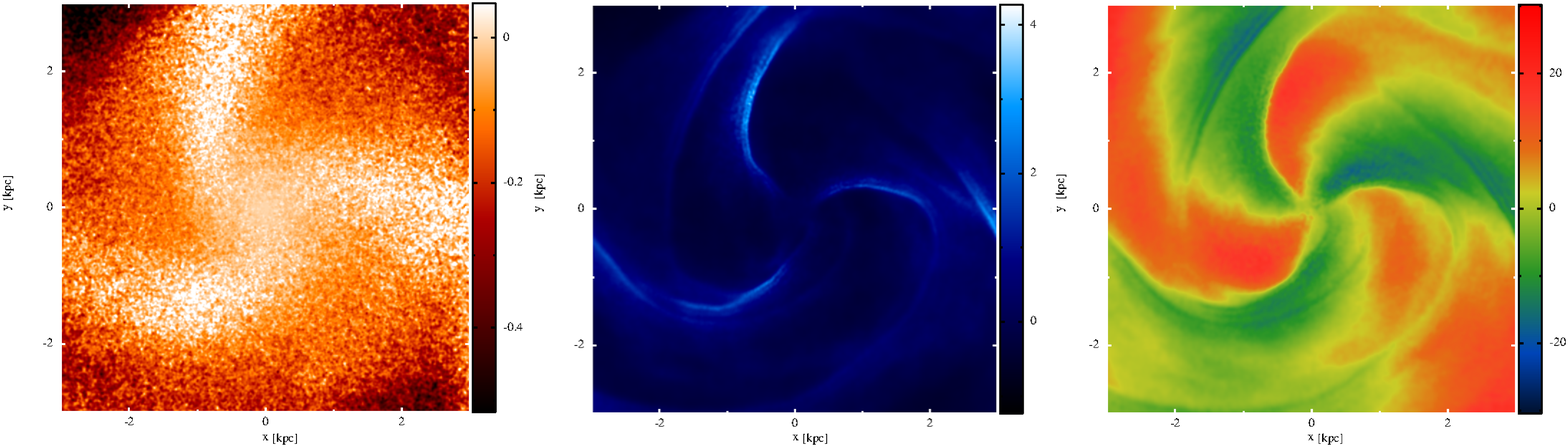}
                \includegraphics[width=19.cm,clip=true,trim=0.5cm 0 0 0cm]{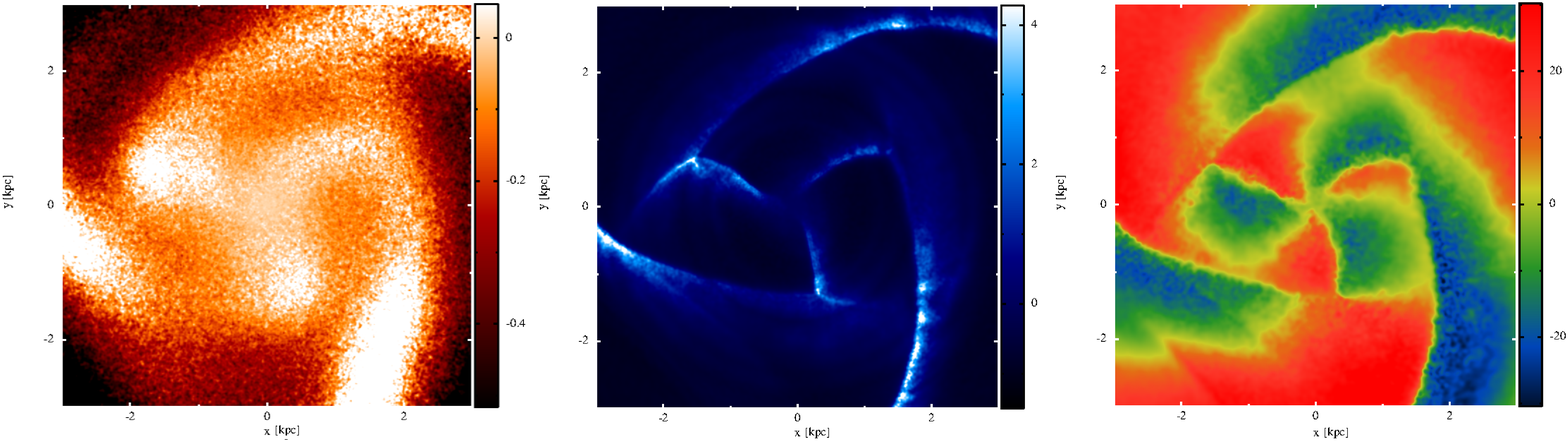}

                \caption{Upper panels: stellar surface density contrast (left), gas
density contrast (middle) and radial velocity map (right, in unit of $\rm km$ $\rm s^{-1}$) for gas in
the inner $3$ kpc at $t=$0.8 Gyr. Lower panels: same as the upper
panels at $t=1.1$ Gyr. See text for details.} 
\label{fig:figInflow80}
\end{figure*}

The high resolution achieved allows us to
resolve sub-kpc scales, and to investigate the detailed causes of the
nuclear gas inflow through the search of stellar and gaseous nuclear
structures. In this section we focus on two times, just before ($0.8$
Gyr) and right at the beginning ($1.1$ Gyr) of the major gas inflow
event.  The properties of the stellar and gaseous distribution at the
two times are highlighted in the upper and lower panels of
figure~\ref{fig:figInflow80}.

The left panels represent the surface density contrast for the stars
within the inner 3 kpc, defined as:
\begin{equation}
\delta_{\Sigma_{\star}}(R,\phi)=\frac{\Sigma_\star(R,\phi)-\langle
\Sigma_{\star}(R,\phi)\rangle_\phi}{\langle 
\Sigma_{\star}(R,\phi)\rangle_\phi},
\end{equation}
where $R$ and $\phi$ are the radial and azimuthal coordinates on the
disc equatorial plane, $\Sigma_{\star}(R,\phi)$ is the stellar surface density
and $\langle \Sigma_{\star}(R,\phi)\rangle_\phi$ is the average stellar surface
density evaluated
in annuli. The central panels show the gas density contrast 
\begin{equation}
\delta_{\rho,{\rm gas}}(R,\phi)=\frac{\rho_{\rm gas}(R,\phi)-\langle \rho_{\rm
      gas}(R,\phi)\rangle_\phi}{\langle \rho_{\rm gas}(R,\phi)\rangle_\phi},
\end{equation}
evaluated on the disc mid-plane.
The right panels show the intensity of the radial motions in $\rm km$ $\rm s^{-1}$.

Before the major inflow event (upper left panel) a single three arm
spiral structure is visible down to scales of about $300$ pc. The gas
is affected by the stellar non-axisymmetric structure and develops
shocks at the edge of the stellar spirals, as observable in density
contrast map (upper central panel). Clear shock fronts develop in the
gas distribution, the gas dynamics is perturbed and radial motions
are triggered (upper right panel). 

A different picture is present at the triggering of the strong gas inflow
episode (lower panels).  At $t=1.1$ Gyr the inner part (within $\approx 1$
kpc) of the stellar three arm spiral structure decouples from the outer spiral
structure, still evident at large scales ($\sim 3$ kpc), as clearly visible in
the lower left panel.  Such decoupled structure is clearly observable in the
gas density (lower central panel) map.  The interplay between the outer and
inner spiral structure increases the radial velocity of the gas in the central
regions as well as the region participating to the radial inflow (lower right
panel). The effect of the inner spiral decoupling onto the gas is
  reminiscent of the bars-within-bars scenario, originally proposed by
  \cite{Shlosman89}, in its {\it ``stuff-within-stuff''} version
  \citep{Hopkins10}, where the gravitational torques acting onto the gas are
  caused by non-axisymmetric structures not necessarily bar-like.

By the time a clear bar forms, all the gas
affected by the nuclear spirals formed the central nuclear knot.

As a final comment, we stress that the nuclear regions of our galaxy do
  not show any evidence of gaseous clumps, whose migration could, in
  principle, cause the major gas inflow event
  \citep[e.g.][]{Bournaud07,Elmegreen08}. As a
  matter of fact, the further inwards one goes, the less evidence one has for
  clump formation. Such trend is not unexpected in systems with small gas-to-stellar
mass fraction as the one we study here, that remains locally
  stable throughout the whole duration of the runs.

\subsubsection{Nuclear disc}

We devote the last part of our analysis to the structure of the gas
nuclear structure forming during the major inflow event. In particular
we will focus on the gas properties well after the nuclear structure
formed and reached a stable configuration.

\begin{figure*}
        \centering
                \hspace{-1.8cm}
                \includegraphics[width=20.cm,clip=true,trim=1.8cm 0 0 0]{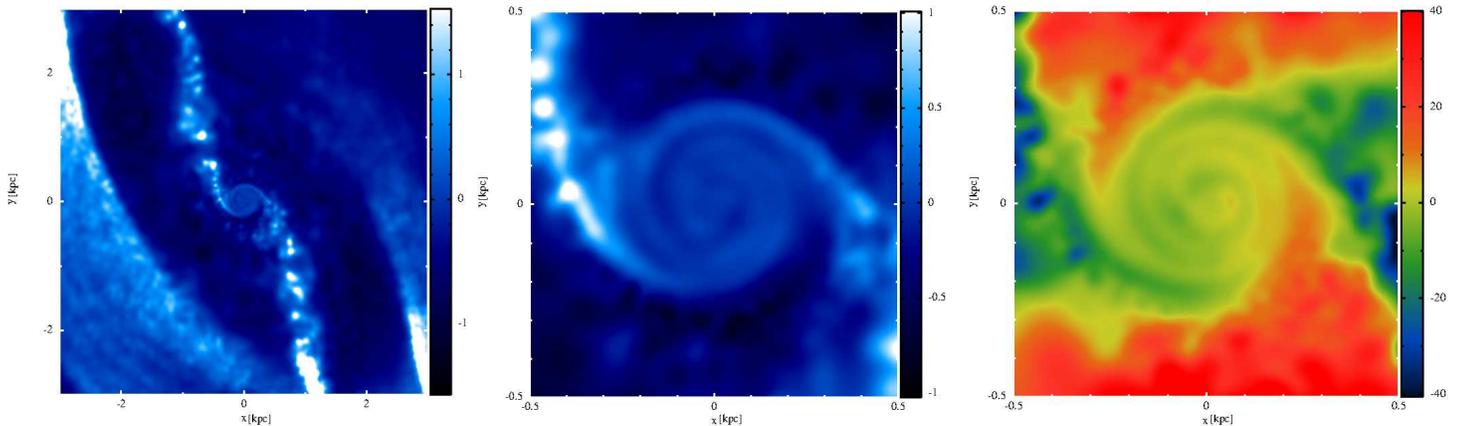}

\caption{Gas density contrast in the inner $3$ kpc (left panel) and 500 pc (central
panel). The right panel shows the radial velocity for gas (in unit of $\rm km$ $\rm s^{-1}$) in the inner 500 pc at 2.5
Gyrs. See text for details.}
\label{fig:inner}
\end{figure*}

Figure~\ref{fig:inner} shows the density contrast of gas in the inner
$3$ kpc (left panel) and in the inner $400$ pc (middle panel), and the
radial velocity map (right panel) at about
$2.5$ Gyr. The orientation of the bar is traced by the inflowing
streams of gas that connect the outer regions of the galaxy with the
inner gaseous structure. The gas in the inner few hundreds of pc forms
a rotating disc. Within the disc nuclear spirals are observable down
to few tens of pc, traced by local gaseous overdensities (lighter
regions in the central panel) corresponding to inflowing gas (blue and green regions
in the right panel). A careful analysis of the stellar
distribution does not show any central structure (neither in the form
of spirals nor of bar). We therefore interpret the central two armed
spirals as the effect that the outer bar has onto the gas within its
ILR, as discussed analytically in \citet{Maciejewsky04a} and observed
in numerical simulations of the response of gas to a bar-like analytical
potential \citep{Maciejewsky04b}.

As a note of caution we stress that the accretion rate at late times
and, more in general, the long-term evolution of the central gaseous
concentration would be significantly different if the gas would be
allowed to form stars, possibly forming a pseudo-bulge like
structure. We plan to study the possible effect of star formation on
the nuclear scale gas dynamics in a future investigation.

\section{Discussion and conclusions} \label{section4}

In this paper we studied the gas response to the formation and evolution of a
stellar bar in an isolated disc galaxy. The galaxy is initially unstable to
the formation of non-axisymmetric structures and develops multiple spiral arms
in the first Gyr, that evolve in a central stellar bar at $t\sim 2$ Gyr. The
forming bar slows down with time, and buckles in its central $\gsim$ kpc
region.

During the first spiral arm dominated phase the gas, forced by the
stellar potential, forms clearly defined spiral arms. During this
phase a major episode of gas inflow takes place, larger by a factor of
$\gsim 3$ than any other inflow event after the bar formation. The analysis of the higher resolution simulation shows that the
  trigger of the major inflow is the decoupling of the nuclear regions
  of the three armed spiral from the outer counterpart. At later
times, when the stellar bar is already established, a low gas density
annulus (here defined as the dead zone) between the bar corotational
and the inner Lindblad resonances $R_{\rm ILR}\lsim R \lsim R_{\rm C}$
is clearly observable in the simulations. We notice that such a gas
depleted region is often observed in local samples of barred spiral
galaxies, as extensively discussed in \citet{Gavazzi15}.

We checked our results against the numerical viscosity used, and we
demonstrated that the gas dynamics is little affected by the exact value of
the viscosity parameter in the SPH runs, and by the exact hydrodynamical
treatment of the gas. We also studied the dependence of our results on the
numerical resolution.  We found that, although the qualitative evolution of
the gas is resolution independent, the exact time at which the non
axisymmetric structures develop and the actual maximum inflow rate at small
(but completely resolved) scales do depend on the resolution achieved. As
discussed above, the difference in the timescales for the inflow and for the
bar formation are probably due to a lower shot noise in the highest resolution
initial conditions. The difference in the magnitude of the maximum inflow
rate, instead, is due to the fact that the bar itself as well as all the
non-axisymmetric structure are better resolved in the highest resolution run,
resulting in a more effective torquing of the gas. 

Independently of the exact numerical implementation, we find that the flux of
gas reaching the most central regions of the galaxy peaks during the bar
formation phase, and not when the bar is fully established. The explanation of
such result is twofold: (1) since bars are quite efficient in driving the gas
within their corotational radius toward the centre, after few bar orbits the
central region of the galaxy is mostly gas free \citep[as already noted
  by][]{Berentzen98}, and there is no remaining gas to be torqued by the bar;
(2) in our simulations the forming bar slows down as the galaxy evolves,
increasing its ILR and corotational radii \citep[in agreement with,
  e.g.][]{Sellwood81, Combes81, Halle15}. As a consequence, the gas that is
perturbed by the early fast-precessing bar reaches regions significantly more
nuclear than gas perturbed at later times.

The low efficiency of large, long-lived and easy to spot bars in fueling the
very central regions of galaxies can explain why many observational studies do
not find significant links between bars and AGN activity. The high efficiency
of the {\it bar formation} process in driving strong inflows toward the very
central region of galaxies hints, on the other hand, at a possibly
underestimated importance of bar driven AGN activity in disc galaxies.

Finally, the analysis of the long lived nuclear gaseous disc
  shows that the outer, large scale, bar keeps on exerting a dynamical
  effect onto the gas within few hundreds of pc. A two armed spiral
  can be observed both in the gas density distribution and in the gas
  dynamics, as already discussed by \citet{Maciejewsky04a} and \citet{Maciejewsky04b}.

As a final note of caution, we highlight the main shortcoming of the
simulation suite discussed here: 
\begin{enumerate}
\item Our simulations lack physically motivated prescriptions for gas
  radiative cooling, star-formation and any star-formation-related feedback,
  as well as accretion onto a possibly present massive black hole and the
  related AGN feedback.We stressed that the dynamics of the gas in the region
  studied here (down to few hundred pc from the galaxy centre) is not strongly
  affected by the lack of additional physics, as the gas inflow happens on few
  orbital timescales. As a matter of fact, we regard the lack of additional
  physics as a plus in our runs, as it allows to clearly highlight the
  dynamical processes ongoing in the simulation in a controlled system. On the
  other hand, the lack of star formation and related feedbacks does not allow
  us to draw firm conclusions about the long term evolution of the gas at
  small scales, as a significant fraction of the gas could turn into stars on
  the Gyr timescales of the simulations. A follow up set of runs including
  additional physics is currently in preparation;
\item All the simulations discussed here share the same idealized initial
  conditions. We regard this as the main drawback of our study. Because of
  this we cannot use our runs to make any general prediction about barred
  galaxies in general. We stress, however, that our simple runs do highlight
  the possible relevance of early gas inflow during the bar formation phase.
  We plan to check our results with fully evolving isolated galaxy
  simulations starting from different initial conditions and, with a
  considerable increase of the computational cost, with cosmological
    simulations.
\end{enumerate}

\section*{Acknowledgments}
We thank the anonymous Referee for her/his suggestions that
  significantly improved the quality of the paper.  We acknoledge
Alessandro Lupi for the help in the technical aspects of the runs, and
for his comments on the paper. We further thank Silvia Bonoli, Pedro
R. Capelo, Guido Consolandi, Jorge Cuadra, Roberto Decarli, and
Giuseppe Gavazzi for their comments and insights.

\bibliographystyle{mn2e}
\bibliography{paper_bars}

\label{lastpage}

\end{document}